\documentclass[english]{article}
\usepackage[T1]{fontenc}
\usepackage[latin9]{inputenc}
\usepackage{graphicx}
\usepackage{esint}

\makeatletter
\newcommand{\lyxaddress}[1]{
	\par {\raggedright #1
	\vspace{1.4em}
	\noindent\par}
}

\makeatother

\usepackage{babel}
\begin{document}
\title{The influence of charge ordering in the microscopic structure of monohydroxy
alcohols}
\author{Martina Požar$^{\ddagger}$, Bernarda Lovrin\v{c}evi\'{c}$^{\ddagger}$,
and Aurélien Perera$^{\dagger}$\thanks{Corresponding author: aup@sorbonne-universite.fr}}
\maketitle

\lyxaddress{$^{\ddagger}$Department of Physics, Faculty of Science, University
of Split, Ru\dj era Boškovi\'{c}a 33, 21000, Split, Croatia.}

\lyxaddress{$^{\dagger}$Laboratoire de Physique Théorique de la Matière Condensée
(UMR CNRS 7600), Sorbonne Université, 4 Place Jussieu, F75252, Paris
cedex 05, France.}
\begin{abstract}
While radiation scattering data provides insight inside the microstructure
of liquids, the Debye relation relating the scattering intensity $I(k)$
to the atom-atom structure factors $S_{ab}(k)$ shows that, ultimately,
it is these individual structure correlation functions which contain
the relevant information about the micro-structure. However, these
quantities are not observables, except in few cases where one can
invert the Debye relation in order to obtain the structure functions.
In the majority of other cases, the need for model dependent computer
simulations is unavoidable. The resulting calculations reveal that
the scattering pre-peak is the result of cancellations between positive
pre-peaks and negative anti-peaks contributions from the atom-atom
structure factors. What of systems where this cancellation is such
that it entirely suppresses the scattering pre-peak? One would be
tempted to falsely conclude that there is no uderlying micro-heterogeneity.
Hence, the structure functions appear as hidden variables, and it
is important to understand the relation between their features and
the micro-structure of the system. Through the computer simulation
study of various mono-ols, ranging from methanol to 1-nonanol, as
well as the branched octanols, we show how the features of the atom-atom
pair correlation function $g_{ab}(r)$ affect that of the structure
factors $S_{ab}(k)$, and reveal that the micro-structure is ultimately
the result of the charge ordering between different atoms in the system.
\end{abstract}

\section{Introduction}

Radiation scattering provides an experimental way of probing the micro-structure
of liquids and mixtures \cite{Textbook_Hansen_McDonald,Textbook_Berne_Pecora,Textbook_Rowlinson_Swinton},
but with disputable insights about the exact nature of this microstructure.
For instance, the intensity $I(k)$ of scattering experiments on colloids
are usually interpreted in terms of the sole colloid structure factor
$S(k)$ using the expression \cite{Textbook_Glatter_scattering}:
\begin{equation}
I(k)=\left[f_{c}(k)\right]^{2}S(k)\label{Ik_coll}
\end{equation}
where $f_{c}(k)$ is the form factor of the colloid, reflecting its
shape, and neglecting all other other molecules present in the solution,
such as the solvent, for example. In the case when the colloid is
a micelle, the proper expression would be the Debye formula \cite{Debye1,Debye2}:
\begin{equation}
I(k)=\sum_{i,j}f_{i}(k)f_{j}(k)S_{ij}(k)\label{IkD_coll}
\end{equation}
where the sum would run over \emph{all pairs of atoms} $i$ and $j$
present in the molecules which make the colloidal solution, and where
the $f_{i}(k)$ are the atoms form factors and $S_{ij}(k)$ the atom-atom
structure factor, including the intra-molecular contributions from
the various molecules in the solution. As to how Eq.(\ref{Ik_coll})
emerges from Eq.(\ref{IkD_coll}) is an open problem. For instance,
one could separate in Eq.(\ref{IkD_coll}) the sum over atoms which
making up the molecules found in the micellar colloids, and would
require to neglect the other atom contributions, namely that of the
solvent, ions and co-ions, as well as the cross-terms between these
molecules and those of the colloidal micelles. It is difficult to
imagine how this could justified, and moreover how the ``shape''
of the micelle would emerge from those terms retained in Eq.(\ref{IkD_coll}).
Yet, it is an experimental fact that both expressions are found equivalent
in many cases. Clearly, in view of the massive amount of data confirming
Eq.(\ref{Ik_coll}), it would be important to show the equivalence
of these 2 equations.

Another similar problem concerns the radiation scattering data from
aqueous mixtures of \emph{small} surfactant like molecules, such as
nonionic polyoxy- ethylene monoalkyl ether surfactants. The experiments
\cite{EXP_SANS_Texeira_Water_DIOLS,EXP_SANS_Texeira_Water_DIOLS_TEMP}
show a variety of behaviour for the $I(k)$ at small $k$, namely
Ornstein-Zernike (OZ) type behaviour and Teubner-Strey (TS) type behaviour.
The former has a large peak at $k=0$, while the latter has a well
defined pre-peak. One might conclude that the corresponding latter
mixtures may have aggregated structures which are responsible for
the pre-peak, while the former may not have such structures. Yet,
computer simulations of aqueous mixtures with simpler alcohols, such
as tertbutyl alcohol, for example, shows micro-domain separation \cite{TBA_VanDerVegt_Lee2005,TBA_Gupta2012,TBA_Bagchi_Banerjee2014,TBA_Overduin2017,AUP_TBA_Overduin2019}.
In this context, it is quite clear that, while various types of aggregative
structure may exist, these may not necessarily lead to the pre-peak
feature, hence posing the question of the exact origin of the presence
or absence of the pre-peak in Eq.(\ref{IkD_coll}).

In the present paper, we study the more modest problem of how the
various atom-atom correlations in neat monools combine to produce
the experimentally observed x-ray scattering intensity, which are
well known to be characterised by the pre-peak feature \cite{EXP_OldAlc,EXP_OldAlc2,EXP_MaginiMethanol,EXP_NartenEthMeth,EXP_FinnsMonools,EXP_MatijaMonools,EXP_Tomsic_butanol,EXP_Nowok_phenyl,2020_Alc_German,2021_octanols_German}. 

\section{Theoretical and simulation details}

\subsection{Charge order and pre-peak}

Charge order in liquids have been documented since the first studied
of ionic melts \cite{Ionic_Hansen1974,Ionic_Hansen1975}, and more
recently in room temperature ionic liquids (RTILs) \cite{Ionic_Triolo2007,Ionic_3,Ionic_1,Ionic_2,Ionic_Caminiti},
particularly in relation to the radiation scattering pre-peak. Model
calculations \cite{AUP_2015} show that the pre-peak, which does not
exist in simple ionic melts, comes from the charge ordering in presence
of uncharged groups attached to them. Alcohols also have such a scattering
pre-peak \cite{EXP_FinnsMonools,EXP_Joarder_Methanol_temp,EXP_JoarderAlcohols,EXP_Karmarkar_TBA,EXP_MaginiMethanol,EXP_MatijaMonools,EXP_NartenEthMeth,EXP_SANS_Soper_pure_methanol,EXP_Sarkar_1propanol,EXP_Silleren_Propanol,EXP_SIM_BenmoreEthanol}.
Since charges in alcohol are tied into the hydroxyl head group, while
these are detached in RTILs into anion and cation groups, it is clear
that charge ordering has nothing to do with polar/apolar order. It
is a very simply the basic Coulomb fact that unlike charges attract
each other while like charges repell. With complex molecules bearing
partial charges, this leads to several near neigbour positioning possibilities,
thus sampling a variety of micro-states. The presence of uncharged
groups introduces a supplementary constraint, since both types of
charged groups try to avoid them. In absence of bonds tying the uncharged
groups to the charged ones, one would quite simply observe a phase
separation, with pre-transitional scattering raise at $k=0$. The
scattering pre-peak can be interpreted as a consequence of the avoided
phase separation, induced by the covalent bonding of the various groups.
In fact, we have demonstrated \cite{AUP_PCCP_prepeak_IL_alc} that
charge order with bond constraints leads to a non-uniform near neighbour
positioning of the charged group, often in form of a chain of oppositely
charges sitting in alternance. This local molecular conformation leads
to typical features in the pair correlation functions $g_{ab}(r)$,
between 2 charged atomic sites of opposite valences, which we have
described elsewhere: the strong pair Coulomb interaction raises the
first contact peak in the $g_{ab}(r)$, while the lack of uniform
spherical distribution of next and higher order neighbours leads to
a depletion correlation feature below one for medium distance in $g_{ab}(r)$.
The first feature gives, by Fourier transform, a wide and high $k=0$
peak in the corresponding structure factor $S_{ab}(k)$ while the
second feature leads to a narrow and small negative anti-peak at $k=0$.
The total contribution to $S_{ab}(k)$ is then a positive pre-peak.
Anti-peaks in $S_{ab}(k)$ happen in a similar way, but for pair of
sites that are weakly correlated or fully uncorrelated, such as a
charged and an uncharged site. For instance, the case of alcohols,
the hydroxyl head group atoms tend to be uncorrelated with the hydrophobic
tail atoms. In this case the first neighbour correlations are depleted,
and mostly below $1$, and raise again above one at medium distances.
This combination leads to a total negative anti-peak, as we will discuss
in the Results section below. 

The total radiation scattered intensity $I(k)$ by molecular liquids
is given by the Debye expression similar to Eq.(\ref{IkD_coll}),
in terms of all the site-site structure factors:
\begin{equation}
I(k)=\alpha\sum_{i,j}\sqrt{x_{i}x_{j}}\sum_{a_{i},b_{j}}f_{a_{i}}(k)f_{b_{j}}(k)F_{a_{i}b_{j}}^{(t)}(k)\label{I(k)}
\end{equation}
where the first sum runs over all the molecular species $i$ and $j$,
and the second sum over all the atoms $a_{i}$ of molecular species
$i$ and $b_{j}$ of molecular species $j$, the $f_{c_{n}}(k)$ are
the form factor for atom type $c_{n}$ of molecular species $n$,
and are radiation dependent, $\alpha$ is a radiation dependent coefficient,
$x_{i}$ is the mole fraction of species $i$, and the total \emph{static}
intermediate scattering function $F_{a_{i}b_{j}}^{(t)}(k)$ (also
called total structure factor $S_{ab}^{(T)}(k)$ in our previous papers
\cite{2020_Alc_German,2019_EthMeth2}) is defined as:
\begin{equation}
F_{a_{i}b_{j}}^{(t)}(k)=F_{a_{i}b_{j}}^{(s)}(k)+F_{a_{i}b_{j}}^{(d)}(k)\label{Ft}
\end{equation}
where the self-part $F_{a_{i}b_{j}}^{(s)}(k)$ is in fact the intra-molecular
structure factor, and the distinct part $F_{a_{i}b_{j}}^{(d)}(k)$
is defined in terms of the atom-atom structure factors $S_{a_{i}b_{j}}(k)$
as
\begin{equation}
F_{ab}^{(d)}(k)=\rho\sqrt{x_{i}x_{j}}S_{a_{i}b_{j}}(k)\label{Fd}
\end{equation}
with
\begin{equation}
S_{a_{i}b_{j}}(k)=\int d\mathbf{r}\exp(i\mathbf{k.r})\left[g_{a_{i}b_{j}}(r)-1\right]\label{Sab}
\end{equation}
The self intermediate scattering function, defined only for atoms
of the same species $i$, has a simple expression for rigid molecules,
in terms of the atom-atom intra-molecular distances $d_{a_{i}b_{i}}$,
as:
\begin{equation}
F_{a_{i}b_{j}}^{(s)}(k)=\delta_{ij}j_{0}(kd_{a_{i}b_{i}})\label{Fs}
\end{equation}
where $\delta_{ij}$ is a Kronecker symbol and $j_{0}(x)=\sin(x)/x$
is the zeroth-order spherical Bessel function. For non-rigid-molecules,
one must compute the self intra-molecular function $g_{a_{i}b_{i}}^{(s)}(r)$
by computer simulations, much the same way one computes the inter
molecular pair distribution function $g_{a_{i}b_{j}}(r),$ by building
the histogram of neighbouring atoms: 
\begin{equation}
g_{ab}(r)=\frac{H_{ab}(r,\delta r)}{N_{i}^{2}\delta V(r,\delta r)}\label{eg-hist}
\end{equation}
where $H_{ab}(r,\delta r$) is the number of atoms $b$ in a spherical
shell of radii $r$ and thickness $\delta r$, centered on atom $a$,
$N_{i}$ is the number of molecules of species $i$ in a volume V,
and $\delta V(r,\delta r)=(4\pi r^{2}\delta r)/V$ is the normalised
volume of the shell.

Eq.\ref{I(k)} shows that positive pre-peaks and negative anti-peaks
can cancel each other, to leave either no peak, or a residual pre-peak
in $I(k)$. It is this residual pre-peak which is interpreted in terms
of the existence of clusters in scattering experiments. However, it
is clear that it is the $g_{ab}(r)$ and the $S_{ab}(k)$ which are
the true indicators of the preferred molecular dispositions inside
the liquid. Interestingly, it is the $g_{ab}(r)$ which is the ultimate
structure indicator, since, as discussed above, the pre-peak of the
structure factor $S_{ab}(k)$ is itself the result of a cancellation
of 2 features of the $g_{ab}(r).$ Therefore, it is perhaps more important
to analyse these functions, rather than the scattering intensity $I(k)$.
Molecular simulation allows to do that, albeit with the price of representing
the true liquids by molecular models, with all the bias that it introduces.

\subsection{Molecular models and simulation details}

The simulations were performed with the program package Gromacs \cite{MD_Gromacs,MD_Gromacs_paralelization},
using the following protocol for all alcohols. The Packmol program
\cite{MD_Packmol} was used to generate random configurations of 2048
molecules for all systems. This system size is sufficient for studying
the structure of monools with long alkyl tails, as shown in our previous
work \cite{2020_Alc_German}. After minimizing the energy, we equilibrated
the systems by doing two consecutive runs of 1 ns, first in the canonical
(NVT) and then in the isobaric-isothermal (NpT) ensemble. The latter
ensemble was used in the following production runs, which lasted 5
ns and sampled at least 2000 congurations for the subsequent calculations
of the site-site correlation functions $g_{ab}(r).$ 

All alcohols were simulated in ambient conditions, at the temperature
of T = 300 K and pressure p = 1 bar. The temperature was maintained
with the Nose-Hoover thermostat (time constant of 0.1 ps) \cite{MD_thermo_Nose,MD_thermo_Hoover},
whereas the pressure was achieved with the Parrinello- Rahman barostat
(time constant of 1 ps) \cite{MD_barostat_Parrinello_Rahman_1,MD_barostat_Parrinello_Rahman_2}.
The time step of 2 fs was used for all systems and the integration
algorithm was leap-frog \cite{MD_INT_leapfrog}. For short-range interactions
the cut-off radius was 1.5 nm. For the long-range Coulomb interactions
we used the particle mesh Ewald (PME) method \cite{MD_PME}, with
FFT grid spacing of 0.12 nm and interpolation order of 4. The LINCS
algorithm \cite{MD_LINCS} was utilized for the constraints.

All of the alcohols were modeled as united-atom, using the classic
OPLS forcefield for alcohols \cite{FF_OPLS_alcohols_1}.

\section{Results}

\subsection{Non-branched mono-ols}

\begin{figure}
\includegraphics[scale=0.45]{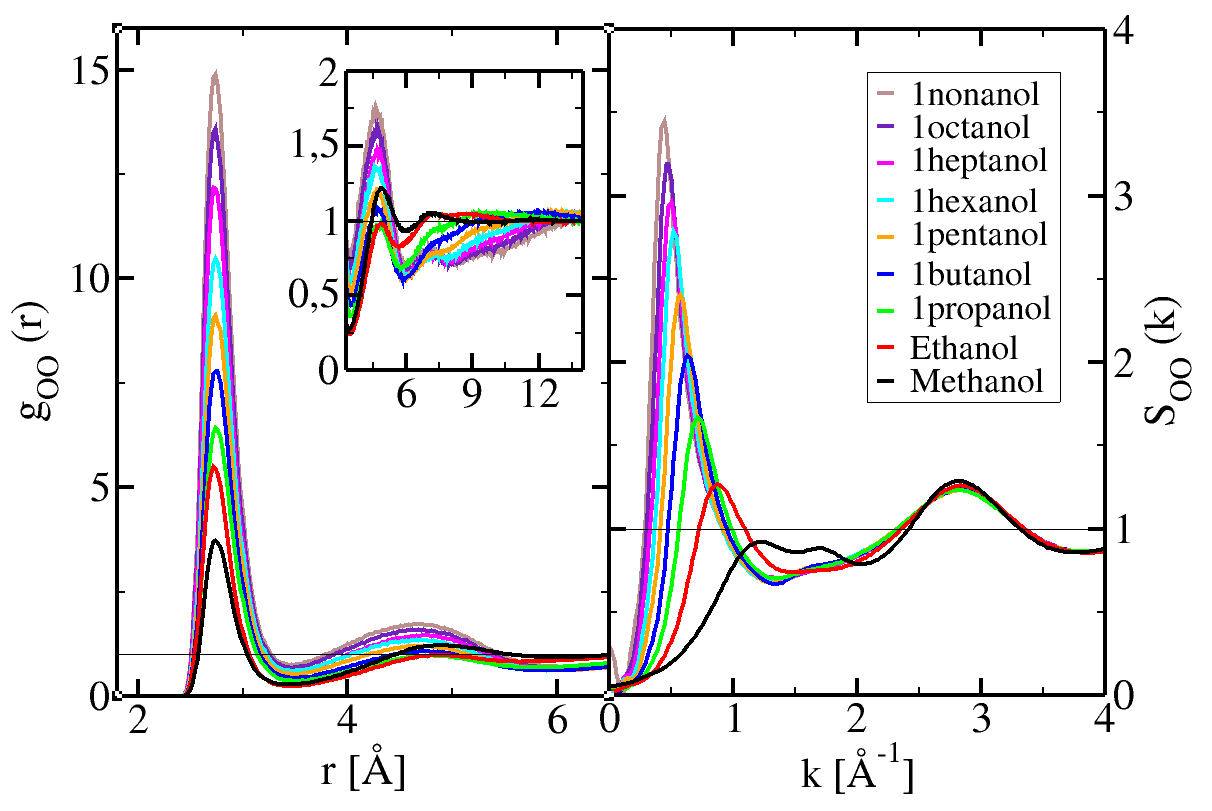}

\caption{Left panel: oxygen-oxygen pair correlation functions $g_{OO}(r)$
for methanol up to 1-nonanol. The inset is a zoom on the depletion
correlation part (see text). Right panel: corresponding structure
factors $S_{OO}(k)$ (the color codes are shown on the side).}

\label{Fig_OO}
\end{figure}

Fig.\ref{Fig_OO} shows the oxygen-oxygen pair correlation functions
$g_{OO}(r)$ in the left panel and the corresponding structure factors
$S_{OO}(k)$ in the right panel, for simulated monohydroxy alcohol
ranging from methanol to 1-nonanol. The left panel shows very cleary
how the O-O correlations increase dramatically from methanol to 1-nonanol,
even though the partial charges on the oxygen and hydrogen sites remain
the same across models, and the contact position remains the same.
The reason for this increase is that, for the same packing fraction
of atoms, there are much less hydroxyl groups in 1-nonanol than in
methanol. Hence, their contact probability increases in inverse proportion
to their rarity, which is just a statistical effect. In contrast,
the depletion part, shown in the inset, increases both in width an
strength when going from methanol to 1-nonanol, suggesting longer
hydroxyl chains in 1-nonanol than in methanol. 

\begin{figure}
\includegraphics[scale=0.45]{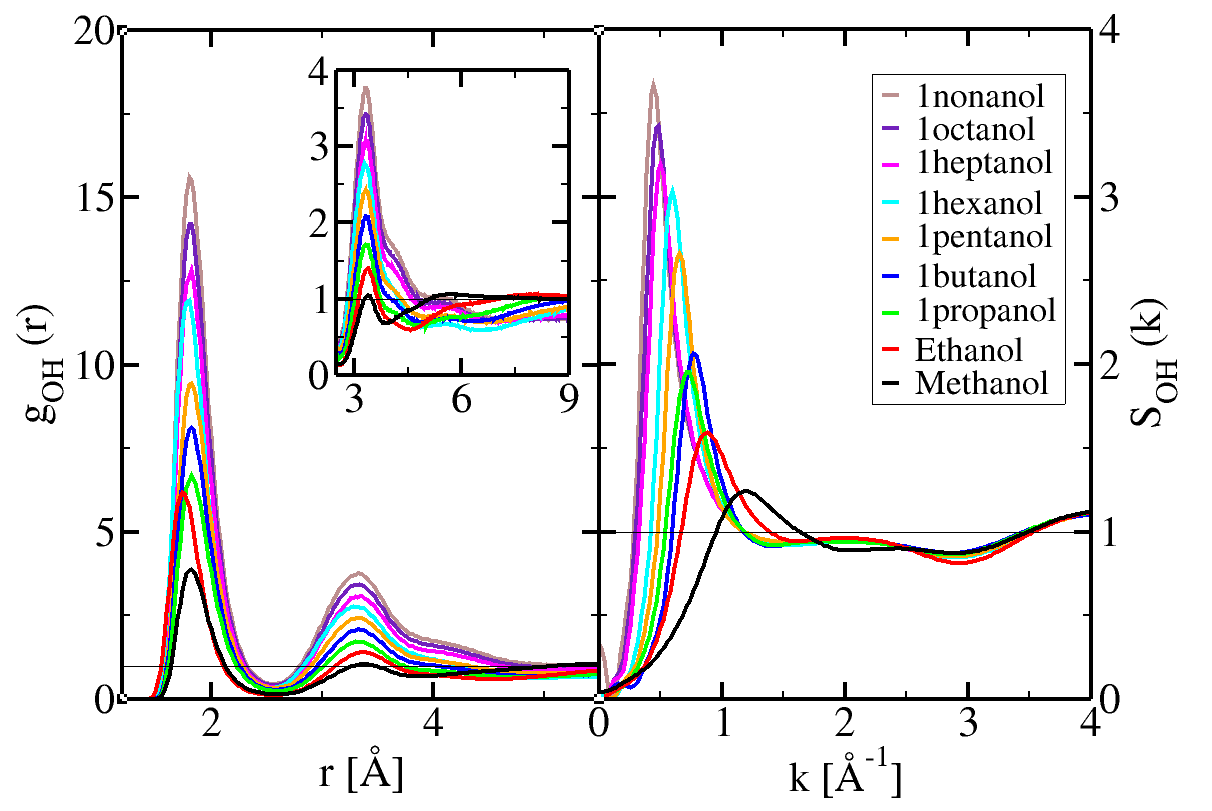}

\caption{Left panel: oxygen-hydrogen pair correlation functions $g_{OH}(r)$
for methanol up to 1-nonanol. The inset is a zoom on the depletion
correlation part (see text). Right panel: corresponding structure
factors $S_{OH}(k)$ (the color codes are shown on the side).}

\label{Fig_OH}
\end{figure}

Both observations are corroborated in the analysis of the structire
factors. The Hbond peak at $k_{HB}\approx3\mathring{A}^{-1}$ is related
to the distance hydrogen bonding distance between oxygen and hydrogen
$r_{HB}=2\pi/k_{HB}\approx2\mathring{A}$) (see Fig.\ref{Fig_OH}
for an explanation). The peak at $k\approx1.5\mathring{A}^{-1}$ ,
which is quite visible for methanol, and barely marked for the other
alcohols, corresponds to the distance $r\approx3.4\mathring{A}$ which
is the diameter of the oxygen atom in the OPLS model \cite{FF_OPLS_alcohols_1}.
The weakness of this peak indicates that the direct, non-bonding,
contact between the two oxygen is not very likely, since it is the
Hbond mediated contact which is preferred. The prominent feature is
the pre-peak, which is at $k_{PP}\approx1.25\mathring{A}$ up to $k_{PP}\approx0.5\mathring{A}^{-1}$
for 1-nonanol, shows the increase of the hydroxyl chains with alcohol
size. This is confirmed by the snapshots from the simulations, which
show clear long chains in the case of longer alcohols. However, the
direct cluster analysis \cite{2020_Alc_German} shows that the\emph{
average} cluster size remain more or less constant across all monohydroxy
alcohols, with about 5 hydroxyl groups per chain. We have no explanation
for this puzzling fact.

Fig.\ref{Fig_OH} shows the oxygen-hydrogen correlations for the same
cases as in previous figure \ref{Fig_OO}. Although the first peak
of the $g_{OH}(r)$ and the pre-peak of the $S_{OH}(k)$ offer strong
similarities with the previous case, we note that the $g_{OH}(r)$
are positioned at the O-H bonding distance, as expected. But the most
important feature is the charge order between like and unlike charges.
This is seen from the peak/anti-peak features which characterise charge
order correlations, as demonstrated in our previous work \cite{AUP_2015,AUP_Tomaz_1}. 

Accordingly, we note that the second neighbour OH peak at $r\approx3.5\mathring{A}$
in Fig.\ref{Fig_OH} coincides with the first minimum of the OO in
Fig.\ref{Fig_OO}. This is even more striking in k-space, when the
peak at $k\approx3\mathring{A}^{-1}$ coincides with the minimum at
the same k-value in Fig.\ref{Fig_OH} right panel. This would be more
apparent if both plots would be superimposed, but aside from this,
it would be also unreadable. These features are simple translation
of the alternation of the O and H atoms along the hydroxyl group chain
aggregates.

\begin{figure}
\includegraphics[scale=0.45]{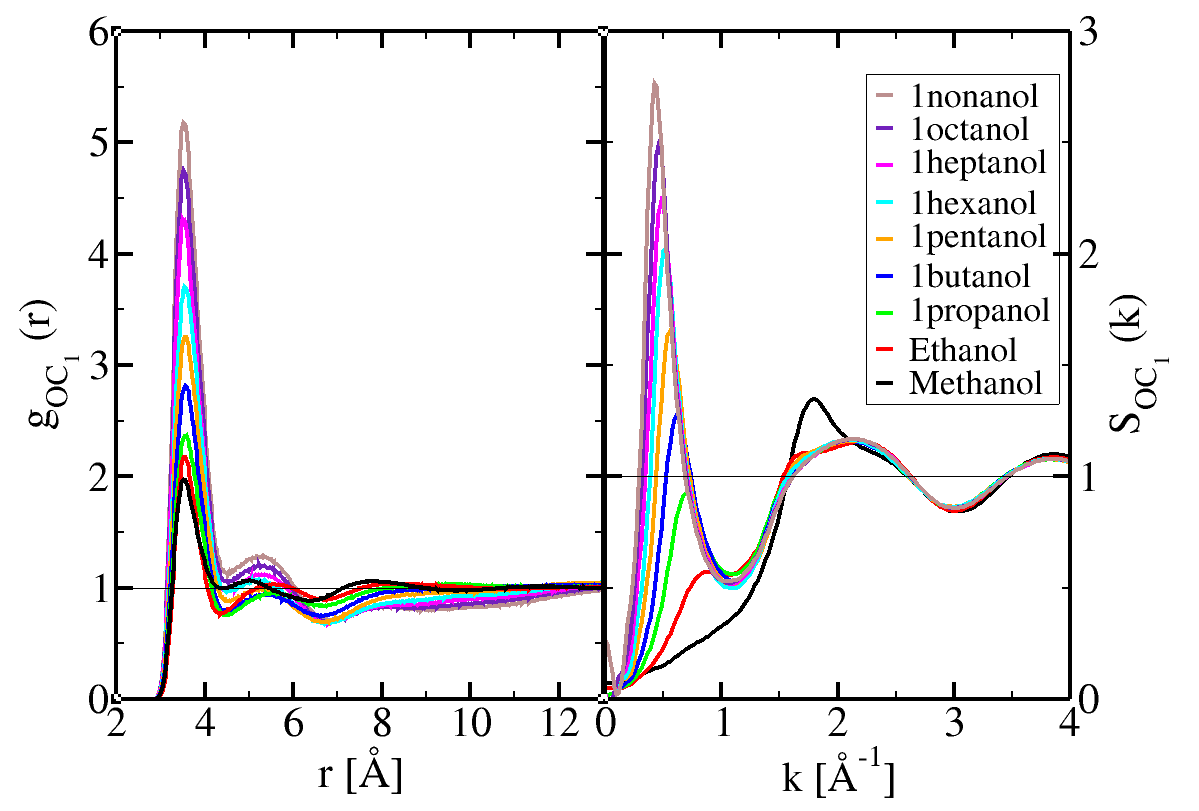}

\caption{Left panel: Pair correlation functions $g_{OC_{1}}(r)$ between the
oxygen and the carbon atom in the first methylene group for methanol
up to 1-nonanol. Right panel: corresponding structure factors $S_{OC_{1}}(k)$
(the color codes are shown on the side).}

\label{Fig_OC_1}
\end{figure}

It is instructive to look at the oxygen atom correlation with the
first carbon of the alkyl chain. In all force field model, this atom
is weakly charged, essentially to neutralize the total charge on the
molecule. The valence is too weak to speak of charge ordering. Nevertheless,
it is important to confirm if this is true. Fig.\ref{Fig_OC_1} shows
the $g_{OC_{1}}(r)$ and $S_{OC_{1}}(k)$. We note that the generic
patterns of the correlations are very similar to that of OO correlations
in Fig.\ref{Fig_OO}, albeit with reduced magnitudes. This is compatible
with the weak charge of the $C_{1}$ carbon. In addition, since this
carbon is the closest to the hydroxyl head group, it seem natural
that the correlations are slaved to that between the oxygens. This
would happen even of the carbon was not charged.

The left panel shows the corresponding structure factors $S_{OC_{1}}(k)$.
We observe again features very similar to that in the left panel of
Fig.\ref{Fig_OO}, albeit with lesser magnitude.

\begin{figure}
\includegraphics[scale=0.45]{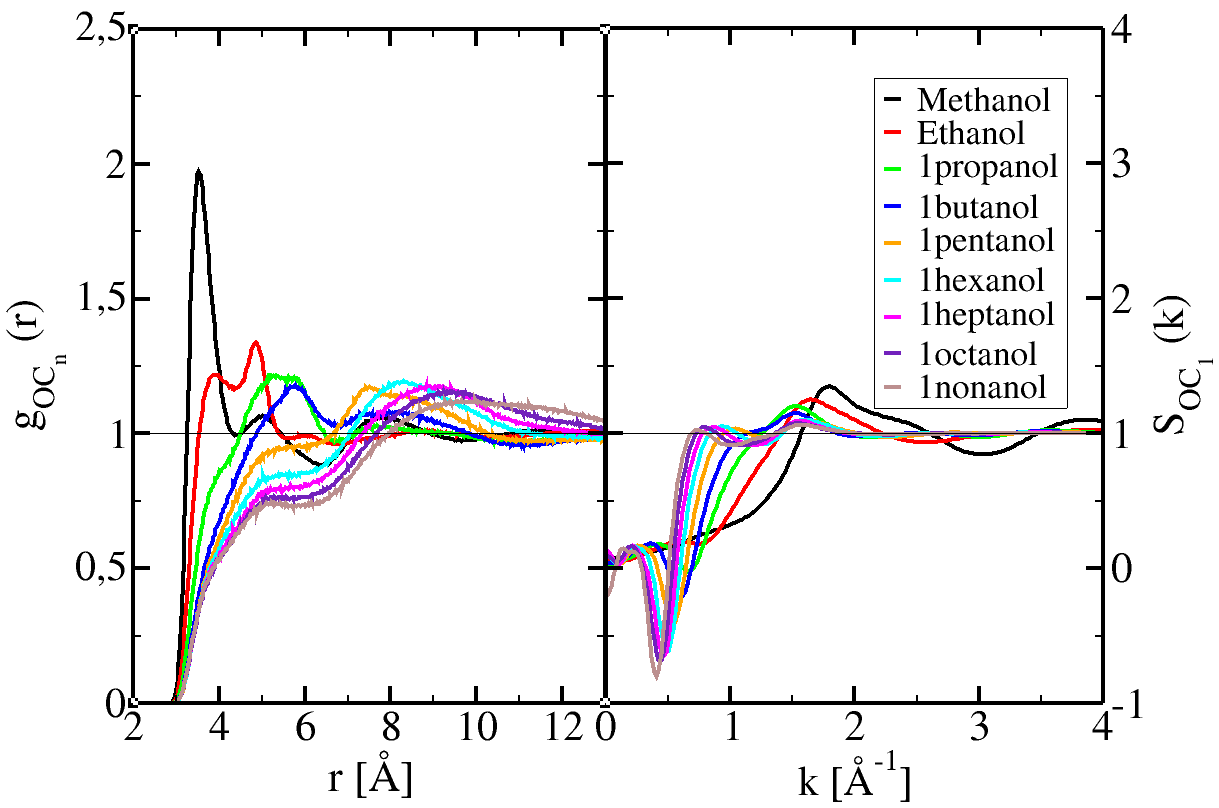}

\caption{Left panel: Pair correlation functions $g_{OC_{n}}(r)$ between the
oxygen and the last methyl group for methanol up to 1-nonanol. Right
panel: corresponding structure factors $S_{OC_{n}}(k)$ (the color
codes are shown on the side).}

\label{Fig_OC_n}
\end{figure}

Now, we look at the inter-molecular correlations between the head
oxygen atom $O$ and the last (uncharged) carbon atom $C_{n}$ of
the methyl group of the alkyl tail. Fig.\ref{Fig_OC_n} shows the
$g_{OC_{n}}(r)$ and $S_{OC_{n}}(k)$. What we observe in the left
panel is that, when the alcohol is short, from methanol until 1-butanol,
there is a dual first neighbour/second neighbour contact high probability,
and the first contact decays rapidly, while the second survives until
1-butanol. Ethanol is interesting since it shows equal probablity
for both contacts. For longer alcohols, the correlations are essentially
depleted. We note that the 3rd correlation peak is in the range $7-9\mathring{A}$,
and is more pronounced for longer alkyl chains than for the shorter
ones. This is somewhat surprizing, since one would imagine exactly
the opposite behaviour. We interpret this as an indication that the
self-assembled objects have better defined shapes and in particular
well defined boundaries for longer alkyl chains. This would be consistent
with the ``inverse line micelle'' picture, with the center of the
cylinder occupied by the hydroxyl chain, and the alkyl tails protruding
all around. The resulting alcohol would then appear as a melt of such
self-assembled line-micelles.

The structure factors in the right panel are a good illustration of
the generic anti-peak description provided in Section 2.1. We observe
the negative anti-peaks, which are the consequence of the depletion
correlations described in the left panel. These are deeper and narrower
for longer alcohols, in proportion of the depth and width of the depletion
correlations.

\begin{figure}
\includegraphics[scale=0.45]{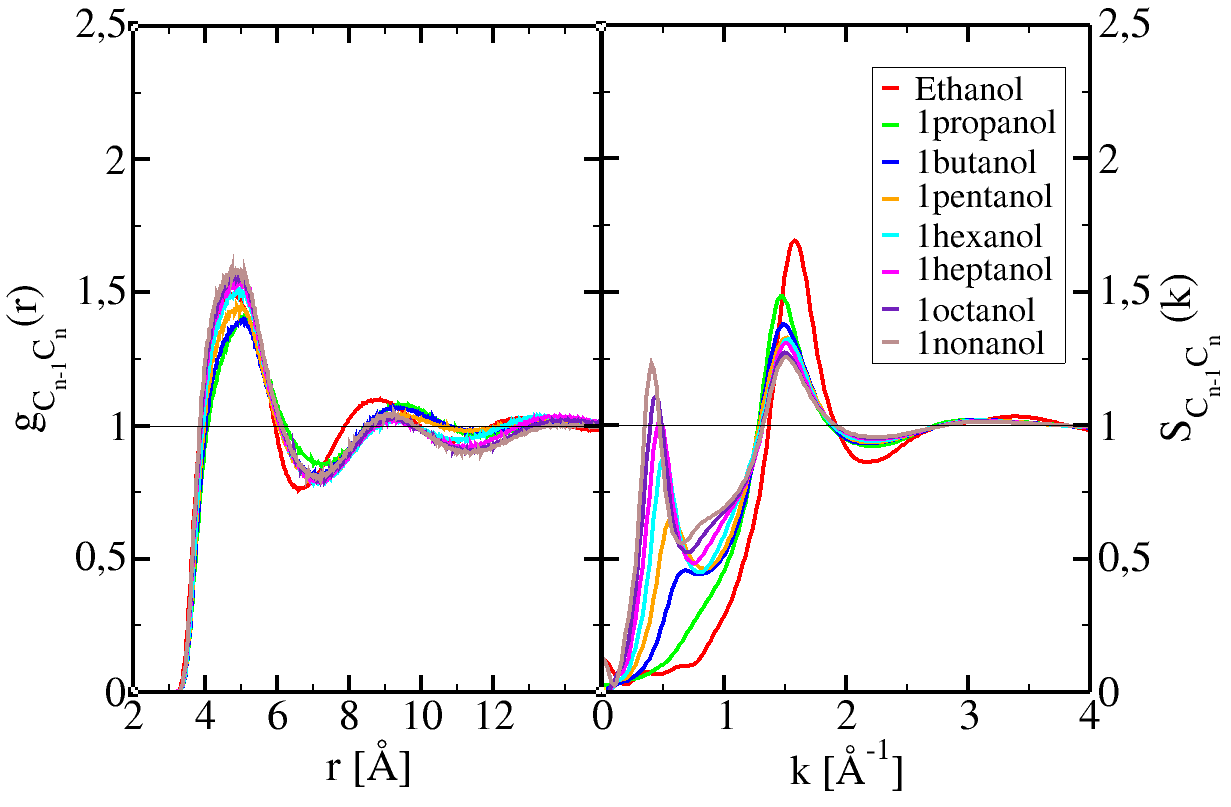}

\caption{Left panel: Pair correlation functions $g_{C_{n-1}C_{n}}(r)$ between
the last methyl group and the one before, for ethanol up to 1-nonanol.
Right panel: corresponding structure factors $S_{C_{n-1}C_{n}}(k)$
(the color codes are shown on the side).}

\label{Fig_CC}
\end{figure}

We turn finally towards the carbon-carbon correlations, and we have
picked those between the last and one before the last carbons of the
alkyl tails. Fig.\ref{Fig_CC} shows the corresponding $g_{C_{n-1}C_{n}}(r)$
and $S_{C_{n-1}C_{n}}(k)$ functions. We observe that these correlations
are strikingly reminiscent of those in a Lennard-Jones liquid for
ethanol and for dimer Lennard-Jones for all others. Since these sites
do not bear any charges, these results appear consistent with the
charge order hypothesis that the main features of the correlations
are slaved to the presence or absence of charge order. However, for
longer alcohols we observe a small depression of high order neighbours. 

This latter trend is further confirmed by the pre-peak in the corresponding
structure factors in the right panel. In fact, the structure factors
bears quite some resemblance with the $S_{OO}(k)$ in Fig.\ref{Fig_OO}
, at the exception that now both the pre-peak and main peak are about
the same height, at least for the longer alcohols above ethanol. This
is very interesting because it witnesses the fact that the tail atoms
share both the simple disorder distribution of Lennard-Jones liquids,
while keeping a memory that they are slaved to the self-assembly of
the head groups. Hence, the role of the tails appears as important
in some aspect. This is further confirmed when looking at branched
alkyl tails, where entropic effects are expected to play an important
role on the self-assembly of the charged head groups.

\subsection{Branched octanolsoctanol}

The n-octanol molecules are a good example to study alkyl tail branching.
The chains are sufficiently long to expect to observe chain packing
entropic effects. The computer simulations reveal that most of the
hydroxyl groups are clustered into chain-like aggregates. Hence, a
good picture of this liquid is that of either a melt of line-like
micelles, or chains ``floating'' in a ``solvent'' of methyl/methylene
super atoms. We expect two types of effects: the influence of the
branching of the tails on the hydroxyl chains formation, and the way
the irrepressible chain formation of the hydroxyl groups affect the
packing of the alkyl tails. Correlation functions are an excellent
observable to answer these questions.

\begin{figure}
\includegraphics[scale=0.45]{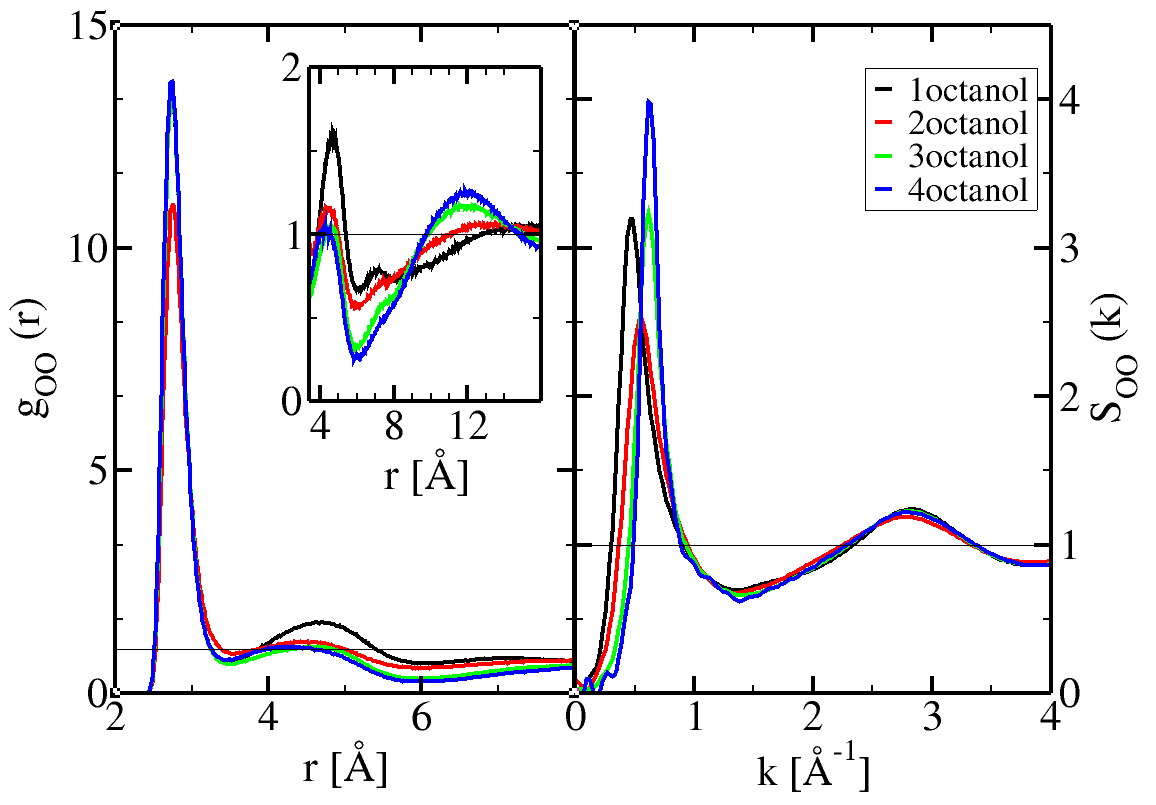}

\caption{Left panel: oxygen-oxygen pair correlation functions $g_{OO}(r)$
for 1-octanol up to 4-octanol. The inset is a zoom on the depletion
correlation part (see text). Right panel: corresponding structure
factors $S_{OO}(k)$ (the color codes are shown on the side).}

\label{Fig_OO-b}
\end{figure}

Fig.\ref{Fig_OO-b} shows the O-O correlations (left panel) and the
corresponding structure factors (right panel). Perhaps the most intringuing
feature is the non-linear behaviour of these curves as a function
of branching. 2-octanol, which is the ``less branched'' of these
octanols, stands out as very different from the others. This is particularly
striking in the first peak of the $g_{OO}(r)$, which is smaller for
2-octanol, while all others coincide almost exactly. In contrast,
both 3-octanol and 4-octanol offer strong similarities (similar depletion
correlations and pre-peak heights). The depletion correlation for
the branched octanols are more marked: the more the branching, the
wider and deeper the depletion. This is a perfect illustration of
the packing entropic problems posed by branching.

We tentatively explain the case of 2-octanol as the difficulty of
aggregating the hydroxyl groups versus the rigidity/flexibility of
the alkyl tail. Indeed, if we consider the idea of a ``sea'' of
methyl/methylene ``solvent'', this poses the problem of packing
these super atoms, while keeping the hydroxyl groups attached together.
The correlation seems to indicate that these problems are maximised
for the case of 2octanol.

\begin{figure}
\includegraphics[scale=0.45]{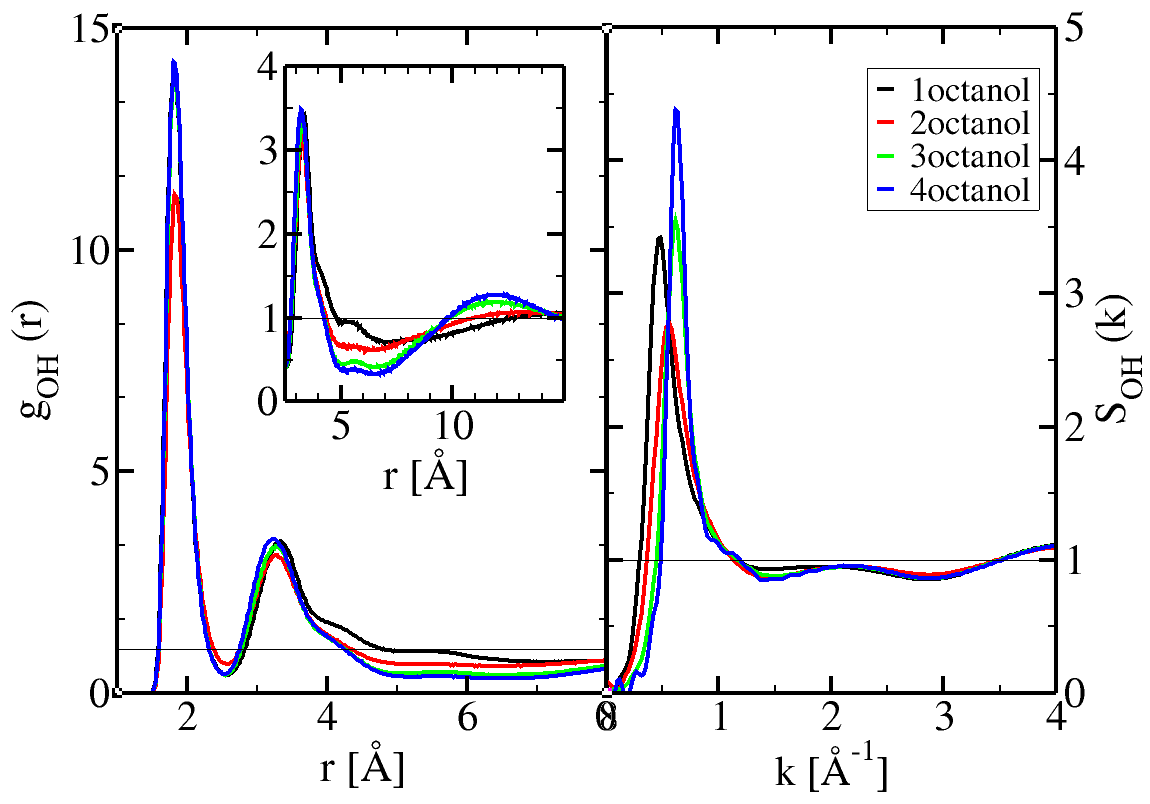}

\caption{Left panel: Pair correlation functions $g_{OH}(r)$ between the oxygen
and the hydrogen for the n-octanol. Right panel: corresponding structure
factors $S_{OH}(k)$ (the color codes are shown on the side).}

\label{Fig_OHb}
\end{figure}

Fig.\ref{Fig_OHb} shows the O-H correlations for the same cases as
in Fig.\ref{Fig_OHb}. The charge order based alternation that we
observed in Figs.\ref{Fig_OO} and \ref{Fig_OH} is more pronounced
with increasing branching. This is directly reflected in the structure
factor pre-peak for 4-octanol being higher than the others. However,
the positions of the pre-peaks change in a non-linear manner: the
k-value decreases from 1-octanol to 3-octanol, and then seems to saturate
since it is nealy the same between 3-octanol and 4-octanol.

\begin{figure}
\includegraphics[scale=0.45]{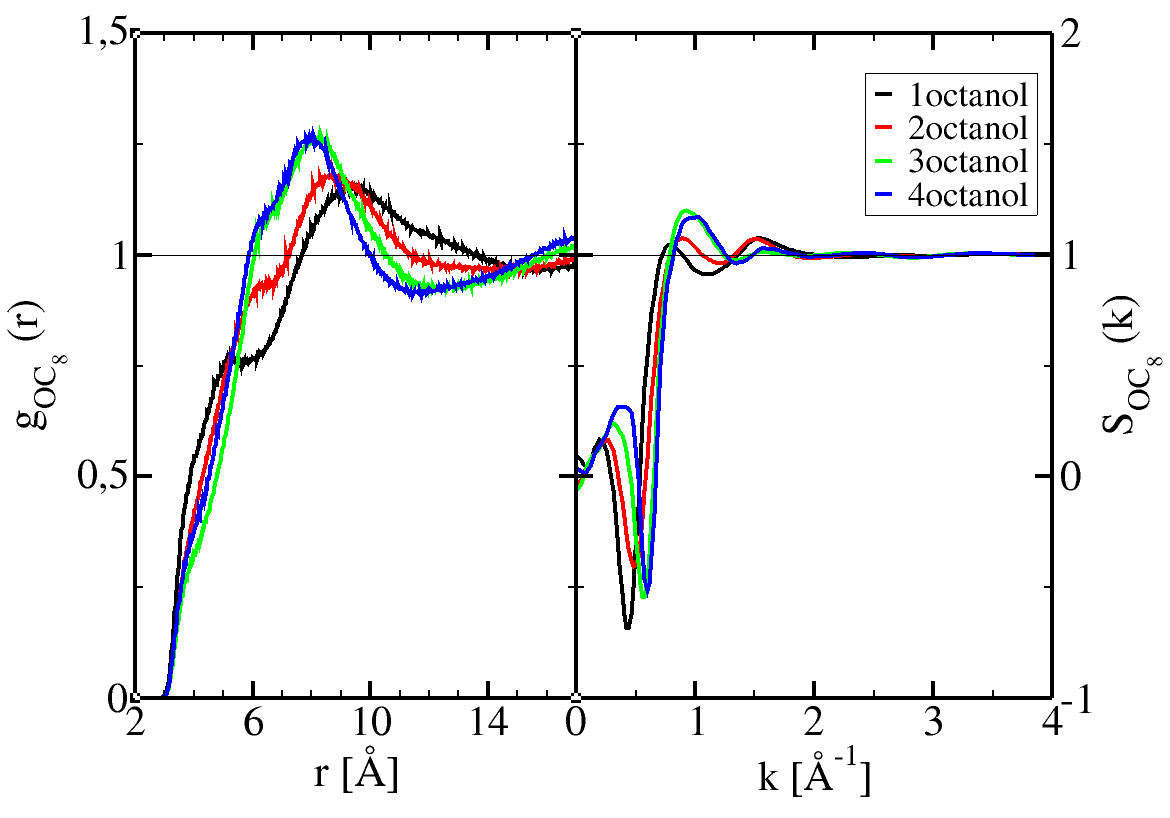}

\caption{Left panel: Pair correlation functions $g_{OC_{n}}(r)$ between the
oxygen and the last methyl group for the n-octanol. Right panel: corresponding
structure factors $S_{OC_{n}}(k)$ (the color codes are shown on the
side).}

\label{Fig_OC_n_b}
\end{figure}

The influence of the tail length and branching can be appreciated
as before by comparing the correlations from the head oxygen to the
last carbon. This is shown in Fig.\ref{Fig_OC_n_b}. We observe that
there are no net pre-peaks. Instead, the shoulder structure at small
$r$ value $r\approx5\mathring{A}$ for 1-octanol, tends to disappear,
while the correlation heights increase from 1-octanol to 4-octanol,
suggesting that these correlations are quite Lennard-Jones like. We
tentatively interpret this trend as as randomisation of correlations
between the oxygen atom and the tail atoms because of increased branching
order. We observe again this saturation of curves at 3-octanol and
4-octanol, since they appear quite similar. 

.

The structure factors show well marked anti-peaks, but of smaller
magnitude than those observed for the shorter and medium lenght non-branched
monools in Fig.\ref{Fig_OH}. Basically, these results reflect the
decorrelation between head groups and tail atoms.

\begin{figure}
\includegraphics[scale=0.45]{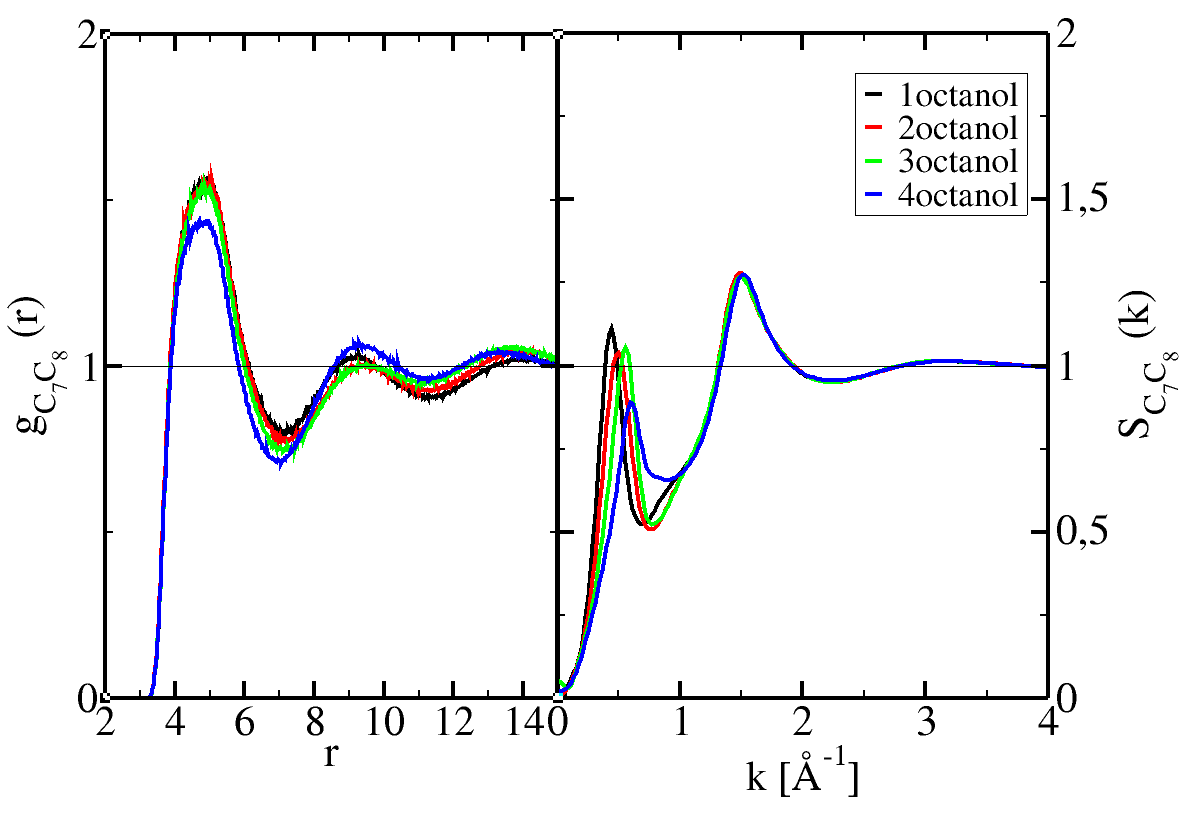}

\caption{Left panel: Pair correlation functions $g_{C_{n-1}C_{n}}(r)$ between
the last methyl group and the one before, for the n-octanol. Right
panel: corresponding structure factors $S_{C_{n-1}C_{n}}(k)$ (the
color codes are shown on the side).}

\label{Fig_CC-1-b}
\end{figure}

Fig.\ref{Fig_CC-1-b} shows the $g_{C_{n-1}C_{n}}(r)$ and $S_{C_{n-1}C_{n}}(k)$
in order to appreciate the correlations between the tail atoms that
the farthest from the charge order constraints. The figures show quite
clearly that the correlations are both Lennard-Jones like and depleted
in the $r$-range $8-14\mathring{A}$. This is confirmed by a net
pre-peak in the corresponding structure factors.

If we recoup these observations with that of the previous figure,
we are forced to conclude that, while head and tail atoms appear randomized,
the tail atoms themselves are still constrained by the charge order
imposed by the head atoms. This is not very surprizing, but it shows
that ``internal'' correlations between aggregated atoms are different
from those outside. This enforces the picture of a coherent self-assembled
object. We can conjecture that self-assembled objects with better
shape visibility, such as micelles will show similar change of correlations
between atoms belonging to the ``inside'' of the object and those
outside. 

\section{Calculated scattering intensities }

The X-ray scattering experiments provide a biased information on the
microscopic structure of the complex disorder liquids, such as those
with aggregates. This is because of the cancelling contributions from
the various structure factors in the k-range $0<k<k_{max}$, where
$k_{max}\approx2\pi/<\sigma>$ corresponds to the mean atomic diameter
size $<\sigma>$ of all the atoms in the molecular system, and which
corresponds also to the main peak in the wide angle x-ray scattering
experiments.

\begin{figure}
\includegraphics[scale=0.45]{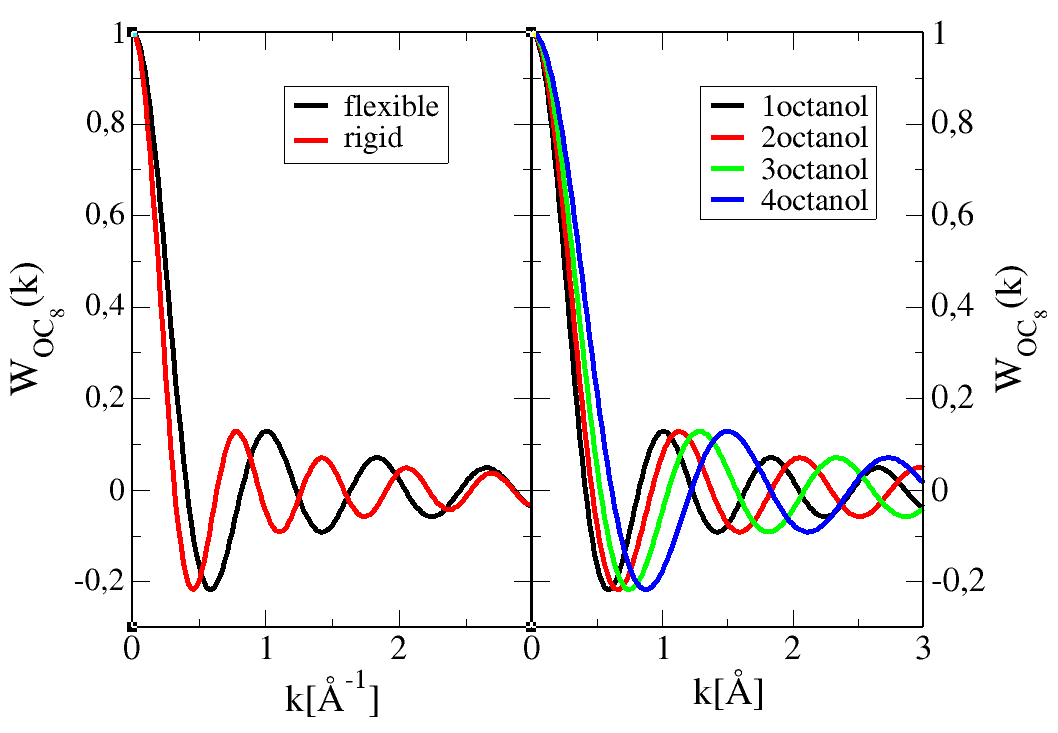}

\caption{Intra-molecular correlations $W_{OC_{8}}(k)$ between the head group
oxygen atom $O$ and last carbon atom $C_{8}$ of the tail. Left panel:
$W_{OC_{8}}(k)$ for 1-octanol; black line for flexible model and
red line for rigid model (see text). Right panel: influence of chain
branching on $W_{OC_{8}}(k)$ for the branched octanol.}

\label{Fig_Wk}
\end{figure}

X-ray spectra calculated from simulations using Eq.(\ref{I(k)}) are
also biased for several reasons. First of all, the model dependence
is strong, as illustrated in our previous studies \cite{2020_Alc_German,2021_octanols_German}.
Secondly, The intra-molecular correlations $W_{ab}(k)$are an important
ingredient since they affect the small-$k$ region. Fig.\ref{Fig_Wk}
illustrates this. The left panel shows a comparison the $W_{OC_{8}}(k)$
between the oxygen $O$ and last carbon atom$C_{8}$ in the case of
1-octanol, between the rigid model (in red) and that (in black) obtained
by explicit calculation in the course of the simulation using the
flexible OPLS model. We see that there is quite a strong dephasing
between the 2 calculations. 

In case of long molecules, the rigid model approximation becomes a
problem. To avoid that, we sample the mean atom-atom distance $d_{ab}$,
as obtained from the simulations, and use the rigid atom approximation
with $W_{ab}(k)=j_{0}(kd_{ab}).$ The right panel shows the intra-molecular
correlations $W_{OC_{8}}(k)$ of the flexible models of various branched
octanols. A larger dephasing corresponds to a shorted distance $d_{OC_{8}}$,
which is consistent in what concerns 1-octanol, but is less trivial
for the others.

\begin{figure}
\includegraphics[scale=0.45]{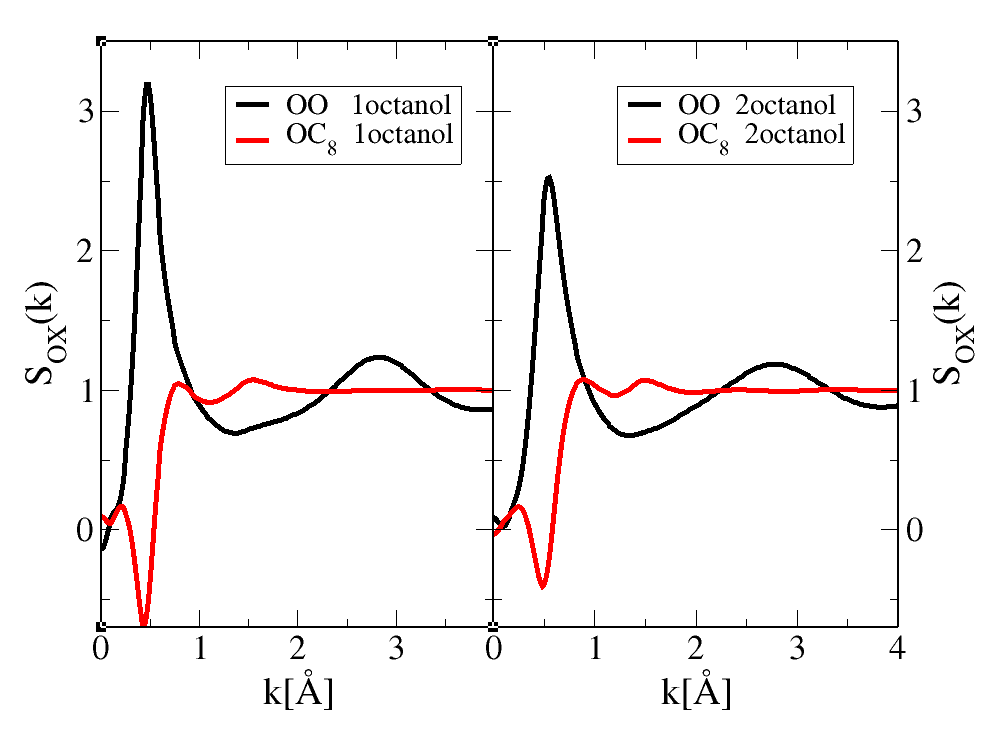}

\caption{Charge order signature on partial structure factors for like-atoms
($OO$) and cross-atom ($OC_{8}$) correlations. Left panel: 1octanol;
right panel: 2octanol.}

\label{Fig_Sk-CO}
\end{figure}

Fig.\ref{Fig_Sk-CO} illustrates the charge order induced pre-peak
anti-peak compensations for 1 and 2-octanol. It shows that the anti-peak
does not quite compensate the pre-peak, in both cases, which is the
reason why a net pre-peak would appear in the scattering intensity
$I(k)$. But, there are cases where these peaks exactly compensate,
and this happen for more spherical aggregative shapes, where the depletion
gains more depth.

\begin{figure}
\includegraphics[scale=0.45]{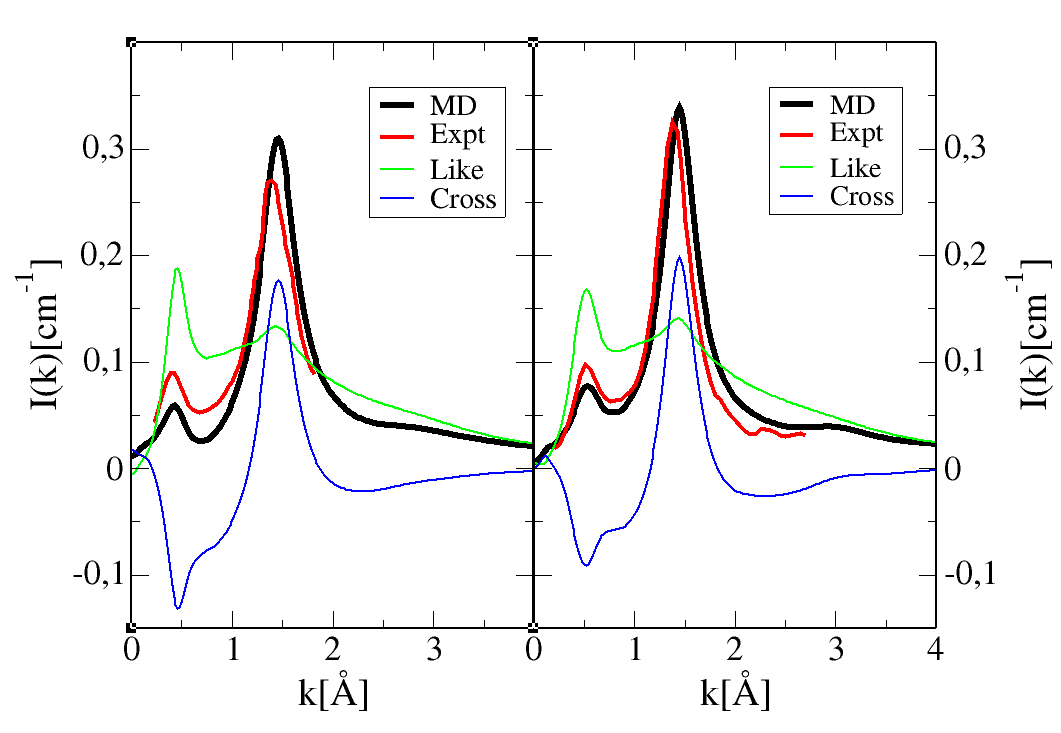}

\caption{x-ray scattering intensities for 1octanol (left panel) and 2octanol
(right panel). Red lines: experimental data from \cite{2020_Alc_German,2021_octanols_German},
black lines: calculated spectra; green lines: total like-atoms contributions
to $I(k)$; blue lines: total corss-atoms contributions to $I(k)$.}

\label{Fig_I(k)}
\end{figure}

Fig.\ref{Fig_I(k)} shows the x-ray scattering intensities, from experiments
(red lines) and calculated (black) for 1 and 2-octanol. The net like-atom
(green) and cross-atom (blue) contributions are equally shown. We
see that both contributions are positive for the main peaks ($k_{max}\approx1.5\mathring{A}$),
while they have opposite signs for the pre-peaks ($k_{PP}\approx0.5\mathring{A}$).
These opposting contributions are directly related to those illustrated
in Fig.\ref{Fig_Sk-CO} in the particular case of $O$ and $C_{8}$
atoms.

.

The agreement between the calculated and experimental spectra have
been discussed in our previous work \cite{2020_Alc_German,2021_octanols_German}.
It is mostly qualitative, particularly for the pre-peak part which
is very sensitive to local atomic labile positionings and force field
details. This is less sensitive for the main peak, which is mostly
controled by the Lennard-Jones interactions between sites, hence the
the part of the interaction which is not responsible for labile self-assembly.

\section{Conclusion}

In this study, we have analyzed the local structure of hydrogen bonded
monohydroxy alcohols, both with non-branched and branched alkyl chains,
through the analysis of the pair correlation functions and associated
structure factors, in order to apprehend the interplay between self-assembly
of the hydroxyl groups in the midst of the alkyl chains. We have shown
that the atom-atom pair correlation functions are the ideal observables
of the molecular micro-structure inside liquids, and that this micro-structure
is chiefly ruled by charge order, including non-charged sites. We
have shown that the pre-peaks and anti-peaks in the structure factors
are a consequence of special highly inhomogeneous local molecular
dispositions induced by charge ordering between the different charged
or uncharged atomic groups. What remains unclear at present is relationship
between the peak/antipeak position, heights and width, in relation
to the supra structure of the self-assembly. Perhaps this is not so
well defined for the small self-assembly such as in these mono-ols,
and it becomes more relevant to discuss these issues with higher forms
of self-assembly, such as with spherical and cylindrical micelles,
vesicles, lamella and others. This poses the problem of the interest
of studying in detail precursor self-assembling systems. Our answer
is that such systems are similar to pre-transitional fluctuations
in spinodal and near critical points of lines. While one can focus
solely on well defined self-assembled shapes, such micelles, and underestimate
the influence of atom-level arrangements and correlations, systems
such as those we have studied herein allow to explore precursor self-assembling
systems, and pay attention to molecular details down to atom level.

\bibliographystyle{jpc_title}
\bibliography{alc2}

\end{document}